\documentclass[twocolumn]{webofc}
\usepackage[varg]{txfonts}
\begin{document}

\title{Quadrupole-octupole coupling and the 
onset of octupole collectivity}

\author{
\firstname{Kosuke} \lastname{Nomura}
\inst{1,2}\fnsep\thanks{\email{nomura@sci.hokudai.ac.jp}}}

\institute{
Department of Physics, 
Hokkaido University, Sapporo 060-0810, Japan
\and
Nuclear Reaction Data Center, 
Hokkaido University, Sapporo 060-0810, Japan
}

\abstract{
Octupole deformation and collective excitations 
are studied within the interacting boson model. 
By using the results of the self-consistent 
mean-field calculations with a universal energy 
density functional, the Hamiltonian 
of the interacting $s$, $d$, and $f$ 
boson system is completely determined. 
A global systematic study confirms 
that significant octupole effects are 
present in actinide, lanthanide, 
and rare-earth nuclei corresponding 
to particular nucleon numbers for 
which octupole correlations 
are empirically suggested to be enhanced. 
}

\maketitle

\section{Introduction}
\label{intro}

In addition to the quadrupole deformation, 
the octupole correlations play roles 
in determining nuclear structure. 
The octupole correlations are enhanced 
in those nuclei with 
specific neutron $N$ or/and proton $Z$
numbers, i.e., 34, 56, 88, 134, $\ldots$. 
Search for the octupole deformed
nuclei has been a 
topic of great interest for experimental 
and theoretical studies 
(see, e.g., \cite{butler2016} for a review). 
Observables that characterize the 
octupole collectivity are the low-lying 
negative-parity band, which forms 
an approximate alternating parity 
band with the positive-parity 
ground-state band, and the large 
$E1$ and $E3$ 
transition matrices connecting the bands. 
Experiments using radioactive-ion beams 
have confirmed evidence for a stable 
octupole shape in a number of nuclei 
in light actinide and lanthanide regions. 
Corresponding theoretical calculations 
have been reported by many authors 
using various nuclear 
structure models.

The interacting boson model (IBM) \cite{IBM} 
has been successful 
in describing low-energy collective 
excitations 
in medium-heavy and heavy nuclei over a wide range 
of the nuclear chart. 
The assumption of the IBM is that 
correlated pairs of valence nucleons with 
spin $0^+$ and $2^+$ are represented 
by a $s$ boson, and a $d$ boson, respectively. 
To compute negative-parity states, 
including those arising from the 
octupole correlations, one should  
introduce in the IBM space octupole, 
$f$ (spin $3^-$), bosons. 
Note that the IBM itself 
is a phenomenological model, that is,  
the parameters have to be obtained 
from experiment. 
It should be, therefore, 
founded on the underlying 
multi-nucleon dynamics, 
so that the model Hamiltonian be 
derived from more fundamental nuclear 
structure models 
\cite{OAI,mizusaki1996,nomura2008}. 
In particular, a mapping technique 
has been developed \cite{nomura2008} 
that establishes the link between the IBM 
and the framework of energy 
density functional (EDF) 
for nuclear many-body systems 
\cite{bender2003,niksic2011,robledo2019}. 
This method consists in using the 
results of the self-consistent mean-field (SCMF) 
calculations based on a university EDF 
to completely determine 
the IBM Hamiltonian, which is then used 
to calculate excitation spectra and 
transition properties. 
$f$ bosons have also been considered 
in the mapping procedure 
\cite{nomura2013oct,nomura2014}, which 
has allowed to study 
octupole deformations and collective 
excitations in a number of mass regions 
(see, \cite{nomura2023oct}, for a review).

This contribution presents recent results 
concerning the octupole collectivity, 
that are obtained from the IBM calculations 
based on the nuclear EDF. 
A brief overview of 
the SCMF-to-IBM mapping is given 
in Sec.~\ref{sec:model}. 
A global study 
of octupole collectivity 
in a large number nuclei (Sec.~\ref{sec:global}),  
and applications to a couple of 
more challenging cases that 
also involve effects of shape 
coexistence (Sec.~\ref{sec:recent}) 
are presented. 
The results presented below are based on 
the works originally published in 
Refs.~\cite{nomura2014,nomura2015oct,nomura2020oct,
nomura2021oct-u,nomura2021oct-ba,
nomura2021oct-zn,nomura2022oct,
nomura2022octcm,nomura2023oct}.

\section{Theoretical framework\label{sec:model}}

\begin{figure}
\begin{center}
\includegraphics[width=\linewidth]
{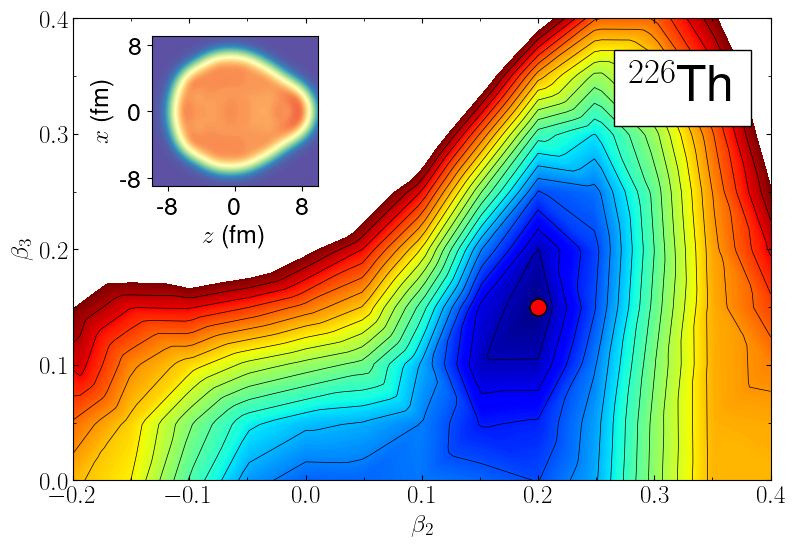}
\end{center}
\caption{\label{fig:pes-th226}
Axially-symmetric quadrupole $\beta_2$ - 
octupole $\beta_3$ SCMF PES, and total intrinsic 
density of $^{226}$Th, computed with the 
constrained RHB method using the 
DD-PC1 EDF and the separable pairing force 
of finite range. Energy difference between 
neighboring contours is 0.5 MeV. 
The red dot stands for the global minimum.}
\end{figure}

Within the EDF framework one often starts with 
the SCMF calculations with constraints on  
multipole moments, e.g., quadrupole, 
octupole, etc., to obtain 
total mean-field energy [hereafter denoted 
as potential energy surface (PES)] that is  
defined in terms of the intrinsic deformation 
variables. 
As an example, Fig.~\ref{fig:pes-th226} 
shows the PES plotted as a function of the 
axially symmetric quadrupole ($\beta_2$) 
and octupole ($\beta_3$) deformations 
for the nucleus $^{226}$Th. 
The PES is here calculated within the constrained 
relativistic Hartree-Bogoliubov (RHB) 
approach employing 
the density-dependent point-coupling 
(DD-PC1) EDF \cite{DDPC1} 
and the separable pairing force 
of finite range \cite{tian2009}. 
One should see from Fig.~\ref{fig:pes-th226} 
a distinct, octupole-deformed 
global minimum at 
($\beta_2,\beta_3$) $\approx$ ($0.2,0.15$). 
Also shown in Fig.~\ref{fig:pes-th226} is the 
projection of the calculated total nucleon 
intrinsic density onto the $x$ and 
$z$ plane of the intrinsic frame,  
corresponding to the $\beta_2$ and $\beta_3$ 
deformations giving the global minimum. 
The density distribution indeed takes 
on a pear-like shape.

The corresponding excited states 
are studied within the IBM. 
Here a version of the 
IBM that comprises 
$s$, $d$, and $f$ bosons (denoted as $sdf$-IBM) 
is considered, with the 
Hamiltonian expressed in general as
\begin{eqnarray}
\label{eq:ham}
\hat H = \hat H_{sd} + \hat H_f + \hat V_{sd-f} \; ,
\end{eqnarray}
where the first, second, and third terms 
on the right-hand side stand for the 
Hamiltonians composed of $s$ and $d$, and of $f$ 
bosons, and the interaction representing 
the coupling between the $sd$- 
and $f$-boson spaces, respectively. 
For each nucleus, 
parameters of the Hamiltonian (\ref{eq:ham}) 
are determined by mapping 
the SCMF PES, $E_{\rm SCMF}(\beta_2,\beta_3)$, 
onto the equivalent one in the IBM, 
$E_{\rm IBM}(\beta_2,\beta_3)$, 
so that the equality 
\begin{eqnarray}
 \label{eq:mapping}
E_{\rm IBM}(\beta_2,\beta_3)
\approx 
E_{\rm SCMF}(\beta_2,\beta_3) \; ,
\end{eqnarray}
is satisfied in the neighborhood of 
the global minimum. 
Note that $E_{\rm IBM}(\beta_2,\beta_3)$ 
is given as the energy expectation value 
taken in the coherent state \cite{ginocchio1980} 
of $s$, $d$, and $f$ bosons. 
The mapped $sdf$-IBM Hamiltonian, with the 
parameters determined by the above 
procedure (\ref{eq:mapping}), is diagonalized 
in the Hilbert space spanned by 
$n(=n_s+n_d+n_f)$ bosons. 
More thorough descriptions of the SCMF-to-IBM 
mapping procedure can be found in 
Refs.~\cite{nomura2014,nomura2023oct}.

\begin{figure}%[h]
\begin{center}
\includegraphics[width=\linewidth]{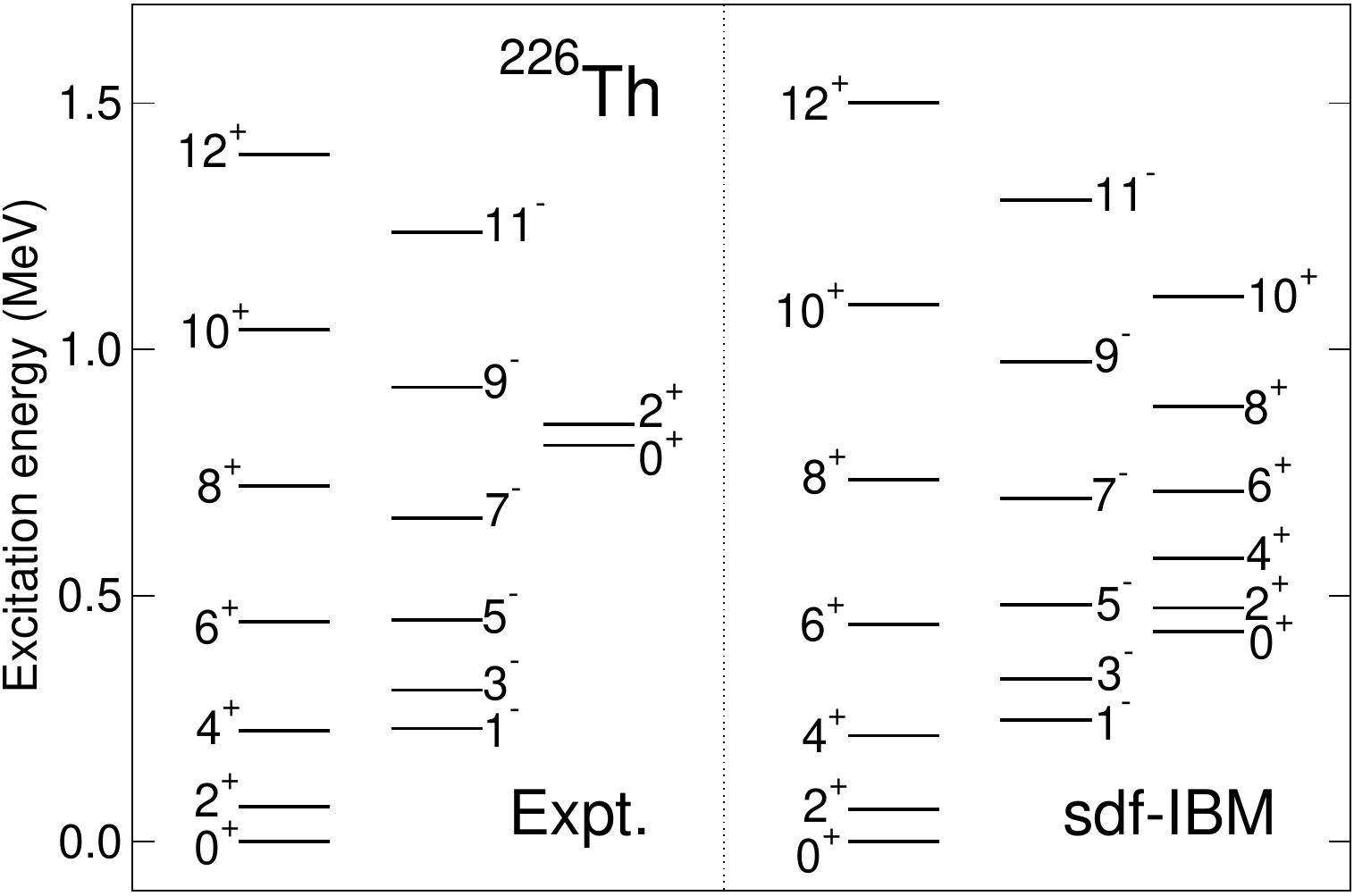}
\end{center}
\caption{\label{fig:th226}
Experimental \cite{data} (left) 
and predicted (right) 
low-energy spectra of $^{226}$Th.}
\end{figure}

Figure~\ref{fig:th226} depicts 
the ground-state band, consisting 
of positive-parity even-spin yrast states, 
and the band of negative-parity 
odd-spin yrast states of $^{226}$Th, 
obtained from the mapped $sdf$-IBM. 
The two bands appear to form a single band 
that resembles the alternating-parity band 
starting with spin $I^\pi \approx 7^-$. 
The excited $0^+_2$ band is also predicted. 
The bandhead, $0^+_2$ state, has been shown 
to be formed by the coupling between two $f$ bosons. 
The $0^+$ excited states of similar kind, 
which are interpreted as being of double octupole 
nature, have been predicted in other 
actinide nuclei \cite{nomura2014}. 
It should be noted that the calculated bands 
from the mapped $sdf$-IBM are in a fairly 
good agreement with experiment, even though 
there is no phenomenological 
adjustment of the IBM parameters to experiment 
and the underlying EDF is not specifically 
tuned to produce octupole deformations.

\section{Global study\label{sec:global}}

The actinide nuclei corresponding to 
$N\approx 134$ and $Z \approx 88$ within 
the mass range $A=200-250$ are among 
the best known for a stable octupole 
deformation and octupole collective 
excitations, the experimental evidence 
being, e.g., 
$^{220}$Rn, $^{224}$Ra, and $^{226}$Th. 
In that context, 
systematic studies on the even-even 
actinide nuclei in the isotopic chains 
from Ra ($Z=88$) to Cf ($Z=98$)
have been performed 
\cite{nomura2020oct,nomura2021oct-u} 
within the mapped $sdf$-IBM with 
the microscopic input provided by the 
SCMF calculations based on the 
Gogny-type EDF. 
The discussion below concerns 
the U isotopes as an illustrative example.

\begin{figure}%[h]
\begin{center}
\includegraphics[width=\linewidth]{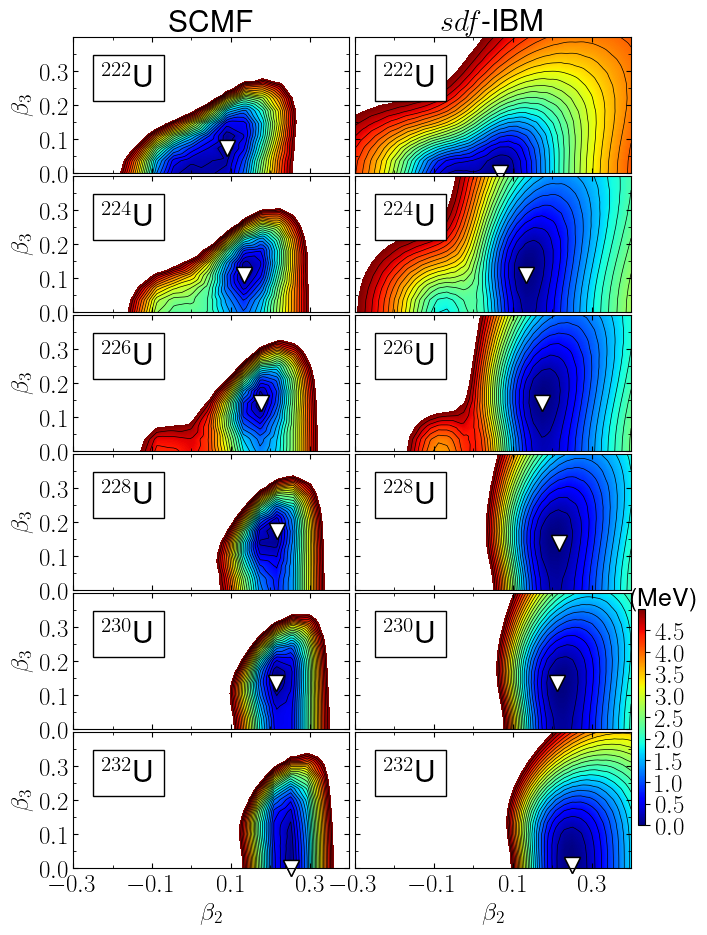}
\end{center}
\caption{\label{fig:pes-u}
(Left column)
$\beta_2-\beta_3$ SCMF PESs 
for the $^{222-232}$U isotopes 
computed by the constrained HFB 
method with the Gogny-D1M EDF. 
(Right column) 
mapped $sdf$-IBM PESs. 
The global minimum is indicated by 
the open triangle.}
\end{figure}

Figure~\ref{fig:pes-u} shows, 
on the left-hand side, 
the SCMF $\beta_2-\beta_3$ PESs for 
the $^{222-232}$U nuclei obtained 
from the constrained 
Hartree-Fock-Bogoliubov (HFB) 
method \cite{robledo2019} 
using the Gogny-D1M interaction \cite{D1M}, 
and the mapped $sdf$-IBM PESs 
on the right. 
While the magnitude of the 
quadrupole deformation remains almost 
constant, the octupole deformation 
appears to develop towards 
$^{226}$U or $^{228}$U, 
at which a pronounced minimum 
at $\beta_3 \approx 0.2$ deformation 
is obtained. The potential, in turn, becomes 
softer in $\beta_3$ direction for the 
nuclei heavier than the $^{228}$U 
isotope, and from $^{232}$U on, an 
octupole deformation is no longer observed. 
The same observation, that the 
nuclear shape evolves from the stable 
octupole to octupole-soft regimes, 
has been shown to apply to the neighboring 
isotopic chains, Ra, Th, Pu, Cm, and Cf 
\cite{nomura2020oct,nomura2021oct-u}. 
Another remark is that, 
comparing between the SCMF and IBM PESs 
for a given nucleus, 
the latter looks flat for larger 
$\beta_2$ and $\beta_3$ deformations. 
This is a general occurrence 
in the SCMF-to-IBM mapping procedure, 
and reflects the fact that the 
IBM has only limited degrees of freedom 
of valence nucleons only, whereas the 
SCMF model consists of all constituent 
nucleons. 

\begin{figure}%[h]
\begin{center}
\includegraphics[width=\linewidth]{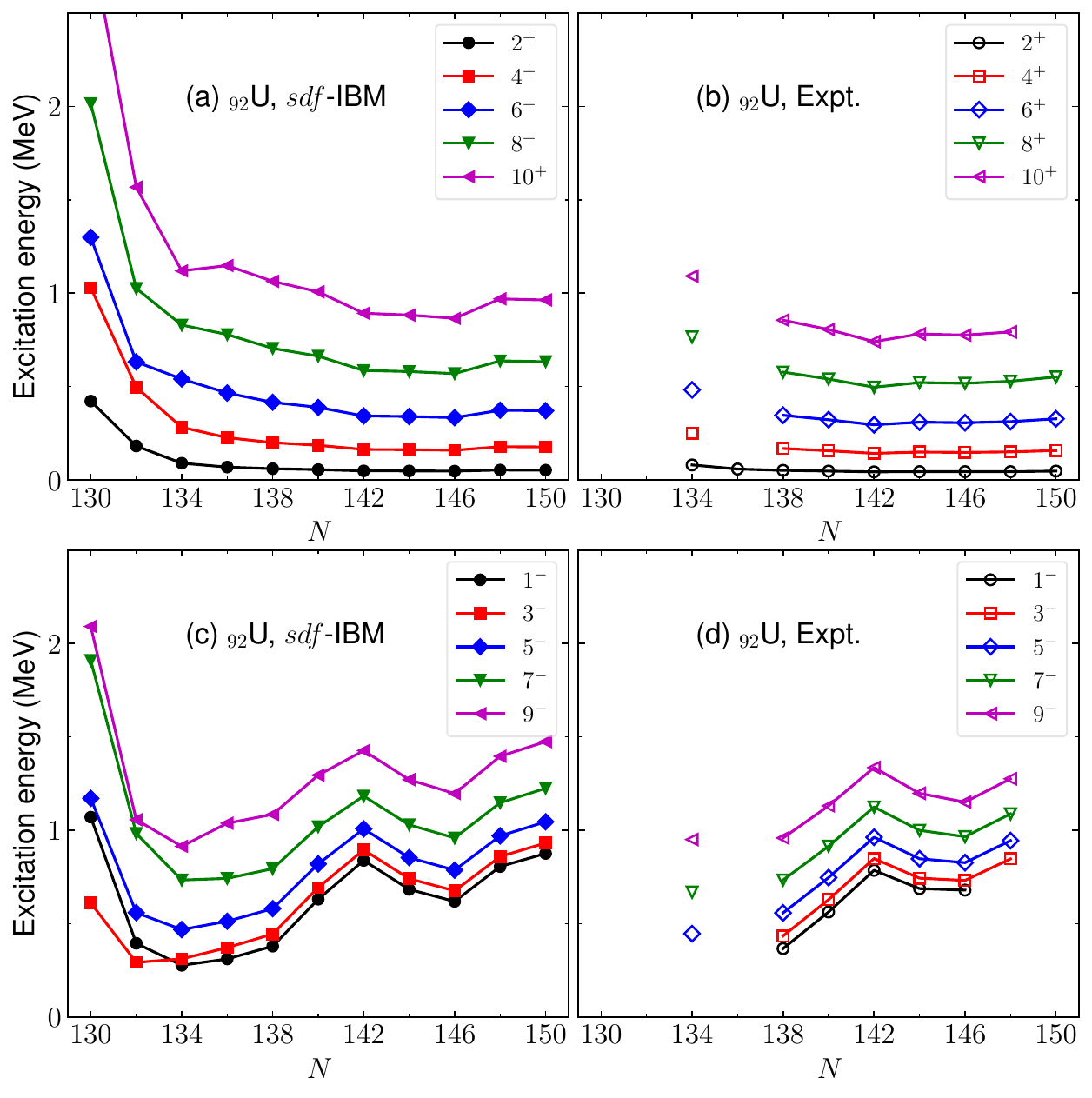}
\end{center}
\caption{\label{fig:u}
Low-energy spectra of the 
positive-parity even-spin (a), 
and negative-parity odd-spin (c) 
yrast states of the 
$^{222-242}$U isotopes 
calculated with the mapped $sdf$-IBM. 
Experimental data \cite{data} 
are shown in (b,d). 
\cite{data}.}
\end{figure}

\begin{figure*}%[b!]
\begin{center}
\includegraphics[width=.49\linewidth]{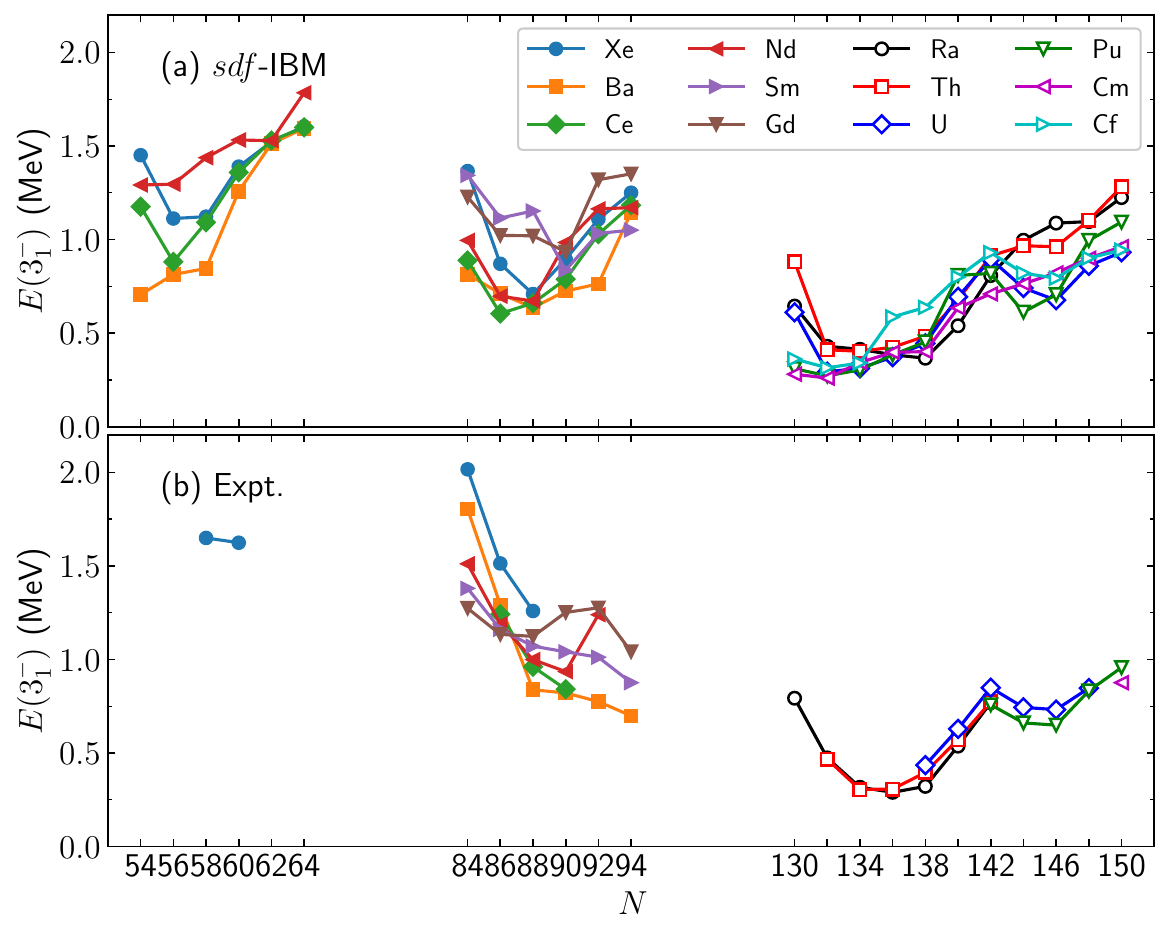}
\includegraphics[width=.49\linewidth]{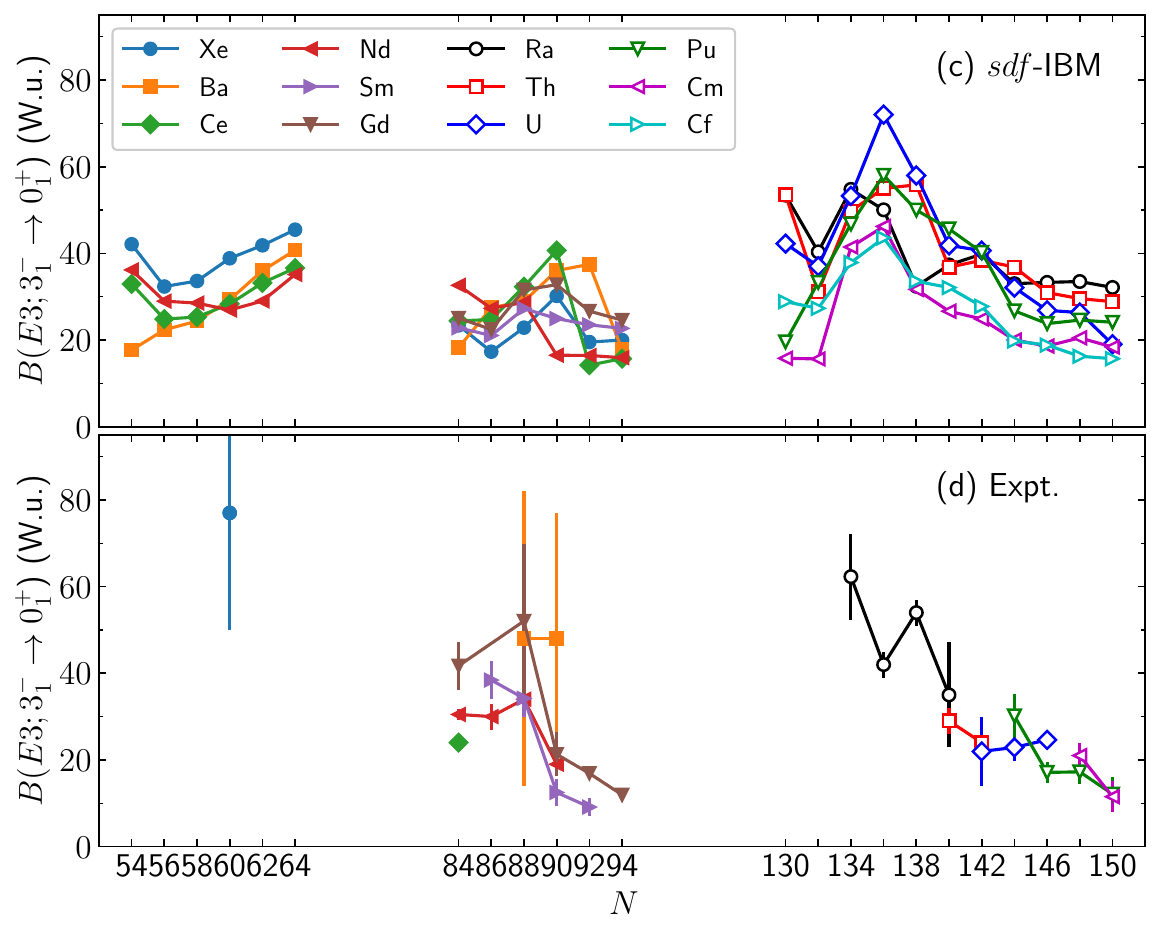}
\end{center}
\caption{\label{fig:3-}
Calculated and observed excitation energies 
of the $3^-_1$ states [panels (a) and (b)], 
and $B(E3;3^-_1 \to 0^+_1)$ 
transition probabilities [panels (c) and (d)]
for a series of isotopes.}
\end{figure*}

Figure~\ref{fig:u} compares the $sdf$-IBM 
and experimental low-lying even-spin 
positive-parity, and odd-spin negative-parity 
yrast states for the U isotopic chain. 
The evolution of the positive-parity 
energy levels shown in Fig.~\ref{fig:u}(a) 
and \ref{fig:u}(b) suggests a rapid 
transition from nearly spherical to 
strongly prolate deformed shapes around 
$N=132$, and one can see a rotational-like 
band structure all the way until $N=150$. 
The calculated negative-parity energy 
levels in Figs.~\ref{fig:u}(c) 
exhibit a parabolic dependence on $N$ 
with minima around $N=134$, at which 
the most pronounced octupole deformed 
ground state is obtained in the PES. 
These negative-parity levels increase 
in energy from $N=134$ to 142, but turns 
to decrease again, showing another 
parabolic behavior around $N=146$, both 
in the calculation and experimentally 
[Fig.~\ref{fig:u}(d)]. 
The appearance of such a local behavior 
may point to the importance of 
dynamical octupole correlations in the 
heavier actinide nuclei around $N=146$. 
This can be also inferred from the fact 
that, even though the PES does not have 
a non-zero $\beta_3$ minimum for those 
nuclei with $N \geqslant 140$, the 
potential is soft in $\beta_3$ deformation, 
hence indicating the considerable amount 
of fluctuation that could 
lower the negative-parity levels.

As well as in the actinide region, 
calculations using the mapped 
$sdf$-IBM framework based on the Gogny-D1M EDF 
have been performed for the isotopes in the 
neutron-deficient \cite{nomura2021oct-zn} 
and neutron-rich \cite{nomura2021oct-ba}
lanthanide, and 
rare-earth \cite{nomura2015oct} regions, 
corresponding to $N/Z$ = 34, 56, and 88, 
in which octupole deformation 
is supposed to be stabilized. 
Figure~\ref{fig:3-} shows the 
computed excitation energies 
of the $3^-_1$ state, $E(3^-_1)$, and 
$B(E3) \equiv B(E3;3^-_1 \to 0^+_1)$ 
transition probabilities in 
comparison to the experimental data.

As one saw in Fig.~\ref{fig:u} 
many of the actinide nuclei with $N=130-150$ 
exhibit a lowering of 
$E(3^-_1)$ around $N=134$. 
The corresponding $B(E3)$ values in 
Fig.~\ref{fig:3-}(c) show an inverse 
parabolic behavior with 
maximal values at $N \approx 134$. 
Another region of interest
is the lanthanide and rare-earth 
nuclei corresponding 
to ($N,Z$) = ($88,56$), from the 
Xe ($Z=54$) up to Gd ($Z=64$) isotopes 
with $N=84-94$. 
The Gogny-HFB SCMF calculations 
in \cite{nomura2015oct,nomura2021oct-ba} 
have produced an octupole deformed 
ground state with small, $\beta_3 \approx 0.1$ 
value, in the PESs for the Xe, Ba, Ce, Nd, 
Sm, and Gd nuclei with $N \approx 88$. 
Similar behavior of the calculated 
$E(3^-_1)$ values is observed 
for these isotopes 
[see Fig.~\ref{fig:3-}(a)]. 
Particularly low-lying $3^-_1$ 
energy levels are obtained 
for the Ba and Ce isotopes. 
The description of the $B(E3)$ values 
in these isotopes is generally good 
[see Figs.~\ref{fig:3-}(c) 
and \ref{fig:3-}(d)].

Also in the neutron-deficient nuclei 
with $N \approx Z \approx 56$, 
the Gogny-D1M SCMF calculation 
in Ref.~\cite{nomura2021oct-zn} 
suggested a few instances exhibiting 
an octupole mean-field minimum: 
$^{112}$Ba, $^{114}$Ba, and $^{114}$Ce. 
As in the other mass regions, 
the $sdf$-IBM produces the 
$E(3^-_1)$ and $B(E3)$ values 
that exhibit parabolic dependencies on $N$. 
However, spectroscopic data on these 
neutron-deficient nuclei is 
very limited in this mass region, 
as it is rather 
close to the proton dripline.

\section{Effects of shape coexistence
\label{sec:recent}}

Octupole correlations should be present 
in those nuclei corresponding to 
($N,Z$) $\approx$ ($34,34$), 
and ($56,34$) as well. 
Description of these nuclear 
systems is, however, even more challenging, 
as the octupole effects are to a large 
extent overshadowed by other 
shape degrees of freedom. 
Neutron-deficient 
$N \approx Z \approx 34$ 
nuclei, for instance, present 
coexistence of prolate and oblate shapes, 
and also the triaxial deformation 
plays a role.  

\begin{figure}%[h]
\begin{center}
\includegraphics[width=\linewidth]
{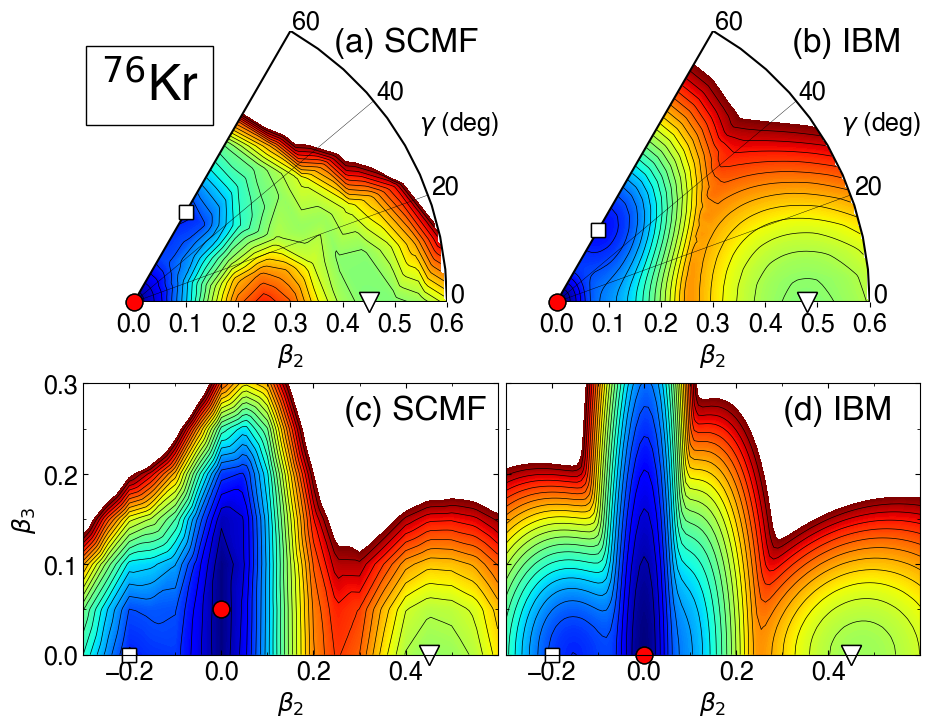}
\end{center}
\caption{
\label{fig:pes-kr76}
Triaxial quadrupole $\beta_2-\gamma$, 
and axially-symmetric $\beta_2-\beta_3$ 
PESs for $^{76}$Kr. 
The SCMF PESs, obtained from 
the constrained 
RHB calculations using 
the DD-PC1 EDF and the separable 
pairing force of finite range 
[panels (a) and (c)], 
are compared with the mapped 
IBM PESs [panels (b) and (d)]. 
Energy difference between neighboring 
contours is 0.1 MeV. 
The spherical global minimum, 
oblate, and prolate local 
minima are represented by the 
solid circles, open 
squares, and triangles, 
respectively. }
\end{figure}

In Ref.~\cite{nomura2022octcm} 
effects of shape 
coexistence in the low-energy quadrupole 
and octupole collective 
states in the neutron-deficient 
$^{72}$Ge, $^{74}$Se, 
and $^{74,76}$Kr nuclei has been 
analyzed. 
For that purpose, the mapped 
$sdf$-IBM has been extended 
to include the configuration 
mixing of normal and intruder states 
which are attributed to the emergence of 
shape coexistence. 
As an example, 
Fig.~\ref{fig:pes-kr76}(a) depicts the 
SCMF PES with the triaxial quadrupole 
$\beta_2-\gamma$ deformations for $^{76}$Kr, 
obtained from the RHB method using the 
DD-PC1 EDF and the separable pairing force. 
There observed three minimum, a spherical 
global minimum, and two local minima 
on the oblate, $\beta_2 \approx 0.25$, 
and on the prolate, $\beta_2 \approx 0.45$, 
sides. 
The $\beta_2-\beta_3$ RHB PES 
is shown in Fig.~\ref{fig:pes-kr76}(c). 
The global minimum appears at finite 
$\beta_3$ deformation, even though 
the potential around the global minimum 
is shallow.

The appearance of the three different 
mean-field minima has been 
incorporated in the $sdf$-IBM 
in the following way. 
First, three independent $sdf$-IBM 
Hamiltonians consisting of $n$, $n+2$, 
and $n+4$ bosons are considered, 
which represent, respectively, 
the normal (0p-0h), and 
two-particle-two-hole (2p-2h), and 
four-particle-four-hole (4p-4h) 
intruder states \cite{duval1981}. 
The normal, 2p-2h, and 4p-4h 
Hamiltonians are then associated with 
the spherical global minimum, 
oblate, and prolate local minima, 
respectively. 
The Hamiltonian of the whole system 
consists of the three Hamiltonian, and 
the interactions that admix different 
configurations, and is diagonalized 
within the extended Hilbert space 
defined as the direct sum of 
the three subspaces. 
Figures~\ref{fig:pes-kr76}(b) and 
\ref{fig:pes-kr76}(d) show, respectively, 
the mapped $sdf$-IBM PESs in 
the $\beta_2-\gamma$, and $\beta_2-\beta_3$ 
planes, with the Hamiltonian completely 
determined by the SCMF calculations. 
For details, see Ref.~\cite{nomura2022octcm}. 

\begin{figure}%[h]
\begin{center}
\includegraphics[width=\linewidth]
{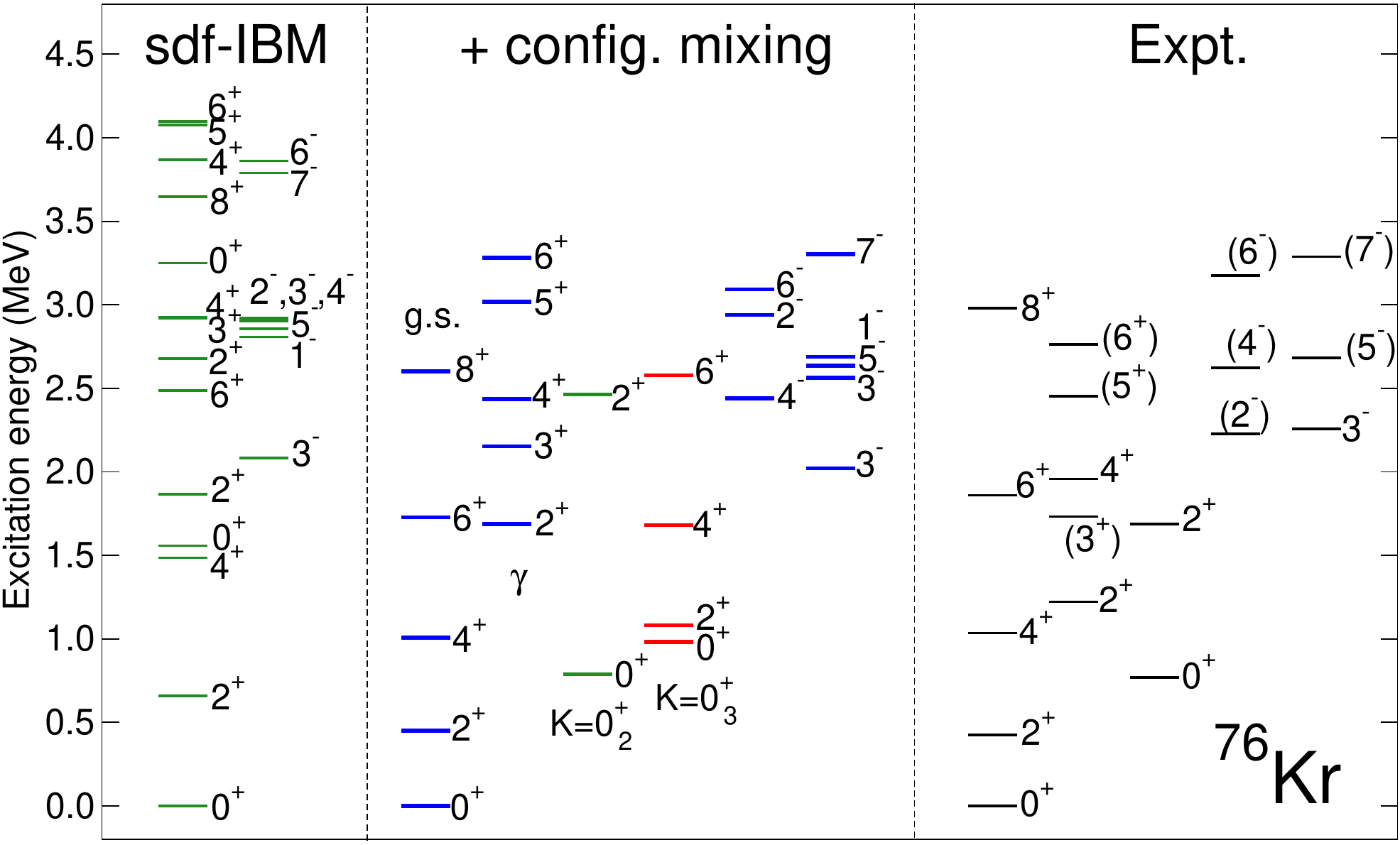}
\end{center}
\caption{\label{fig:kr76}
Calculated low-energy spectra 
for the $^{76}$Kr nucleus obtained 
from the $sdf$-IBM that does not 
(left panel) and does (middle panel) 
include the configuration mixing. 
The observed energy spectrum is 
shown on the right panel.}
\end{figure}

Figure~\ref{fig:kr76} compares 
the low-energy spectra 
for $^{76}$Kr between the $sdf$-IBM 
that includes only the single configuration 
associated with the spherical global 
minimum, and the $sdf$-IBM that takes 
into account the configuration mixing 
of the three boson subspaces associated 
with the spherical, oblate, and prolate 
mean-field minima. 
A significant impact of introducing 
the configuration mixing 
is that the excited $0^+$ 
levels are dramatically lowered. 
The predicted $K^\pi=0^+_2$ and $0^+_3$ 
bands are associated with the 
oblate 2p-2h, and prolate 4p-4h 
configurations, respectively, while 
the ground-state (g.s.), $\gamma$, 
and negative-parity bands are 
of normal, 0p-0h one. 
The description of negative-parity levels 
is also improved by the inclusion 
of configuration mixing. 
Similar results have been obtained for 
neighboring isotopes, 
$^{72}$Ge, $^{74}$Se, and $^{74}$Kr 
\cite{nomura2022octcm}.

The mapped $sdf$-IBM has also been 
applied to the 
neutron-rich $N\approx 56$ Se, Kr, Sr, 
Zr, and Mo isotopes \cite{nomura2022oct}. 
The SCMF calculations 
using the RHB model with the DD-PC1 EDF, 
however, have not produced any 
octupole-deformed ground state 
even at $N=56$. 
The corresponding $sdf$-IBM calculations 
provided the lowest negative-parity states 
at the excitation energies of 
$\approx 1.5 - 2.0$ MeV, but failed in 
reproducing behaviors of positive-parity 
low-spin states in the Zr isotopes 
that exhibits an abrupt decrease of 
energy levels at $N=60$. 
For the neutron-rich nuclei 
in this region, using only 
the axially symmetric 
$\beta_2$ and $\beta_3$ degrees of freedom 
may not be sufficient, and some 
other degree of freedom such as triaxiality 
may need to be considered, or the problem 
may well point to a certain deficiency of 
the underlying EDF.

\section{Summary\label{sec:summary}}

Recent theoretical 
calculations on the quadrupole 
and octupole collective 
states in a large number of medium-heavy 
and heavy nuclei within the $sdf$-IBM that is 
based on the nuclear EDF have been presented. 
A global systematic study 
using this framework has confirmed 
many instances for enhanced 
octupole collectivity 
at those nuclei corresponding to 
the nucleon numbers 34, 56, 88, and 134, 
for which the octupole deformation 
has been observed. 
On the other hand, it has been 
suggested that 
some higher-order effects, 
such as the coexistence of different 
intrinsic shapes, or additional 
shape degrees of freedom, e.g., 
quadrupole triaxial, tetrahedral, 
hexadecapole, etc. deformations, 
need to be incorporated 
in the SCMF-to-IBM mapping procedure
for precise descriptions of 
the proton-rich $N \approx Z \approx 34$, 
and neutron-rich $N \approx 56$ 
nuclei.

\end{document}